\documentclass[twocolumn,showpacs,prbprintnumbers,prb,fleqn,superscriptaddress]{revtex4}
\usepackage{epsfig}
\usepackage{graphicx}
\usepackage{amsfonts}

\newcommand{\ct}{\cite}
\newcommand{\bi}{\bibitem}
\newcommand{\be}{\begin{equation}}
\newcommand{\ee}{\end{equation}}
\newcommand{\ba}{\begin{eqnarray}}
\newcommand{\ea}{\end{eqnarray}}
\newcommand{\al}{\alpha}
\newcommand{\ep}{\epsilon}
\newcommand{\ga}{\gamma}
\newcommand{\non}{\nonumber}

\newcommand{\ket}[1]{|#1\rangle}

\begin{document}
\title{Defect generation in a spin-1/2 transverse $XY$ chain under repeated
quenching of the transverse field }
\author{Victor Mukherjee}
\email{victor@iitk.ac.in}
\author{Amit Dutta}
\email{dutta@iitk.ac.in}
\affiliation{Department of Physics, Indian Institute of Technology Kanpur,
Kanpur 208 016, India} 
\author{Diptiman Sen}
\email{diptiman@cts.iisc.ernet.in}
\affiliation{Center for High Energy Physics, Indian Institute of Science, 
Bangalore 560 012, India}

\begin{abstract}
We study the quenching dynamics of a one-dimensional spin-1/2 $XY$ model in 
a transverse field when the transverse field $h(=t/\tau)$ is quenched 
repeatedly between $-\infty$ and $+\infty$. A single passage from $h \to -
\infty$ to $h \to +\infty$ or the other way around is referred to as a 
half-period of quenching. For an even number of half-periods, the transverse
field is brought back to the initial value of $-\infty$; in the case of 
an odd number of half-periods, the dynamics is stopped at $h \to +\infty$. The
density of defects produced due to the non-adiabatic transitions is calculated
by mapping the many-particle system to an equivalent Landau-Zener problem and 
is generally found to vary as $1/\sqrt{\tau}$ for large $\tau$; however, the 
magnitude is found to depend on the number of half-periods of quenching. For 
two successive half-periods, the defect density is found to decrease in 
comparison to a single half-period, suggesting the existence of a corrective
mechanism in the reverse path. A similar behavior of the density of defects 
and the local entropy is observed for repeated quenching. The defect density 
decays as $1/{\sqrt\tau}$ for large $\tau$ for any number of half-periods, 
and shows a increase in kink density for small $\tau$ for an even number; the 
entropy shows qualitatively the same behavior for any number of half-periods. 
The probability of non-adiabatic transitions and the local entropy saturate to
$1/2$ and $\ln 2$, respectively, for a large number of repeated quenching.
\end{abstract}

\pacs{73.43.Nq, 05.70.Jk, 75.10.Jm}
\maketitle

\section{Introduction}

Zero-temperature quantum phase transitions \ct{sachdev99,dutta96} driven by 
quantum fluctuations have been studied extensively in recent years. In a 
quantum system, statics and dynamics are intermingled and hence a quantum 
critical point is associated with a diverging length scale as well as a 
diverging time scale called the relaxation time. The relaxation time of a 
quantum system is the inverse of the minimum energy gap which goes to zero at 
the quantum critical point. In the proximity of a quantum critical point, the 
spatial correlation length $\xi$ grows as $\xi \sim |\delta|^{-\nu}$ and the
characteristic time scale or the relaxation time $\xi_\tau$ scales with $\xi$ 
as $\xi_\tau \sim \xi^z$, where $\delta$ is the measure of the deviation from 
the critical point, and $\nu$ and $z$ are the critical exponents 
characterizing the universality class of the quantum phase transition.

When a quantum system is swept across a zero-temperature critical point 
\ct{sachdev99,dutta96} by slowly varying a parameter of the Hamiltonian at a 
uniform rate, the resulting dynamics fails to be adiabatic, however slow the 
time variation may be. This is because of the diverging relaxation time 
discussed above. There exist non-adiabatic transitions which eventually lead 
to defects in the final state. According to the Kibble-Zurek (KZ) argument 
\ct{kibble76}, the non-adiabatic effects dominate close to the critical point 
where the rate of change of Hamiltonian is of the order of the relaxation time
of the quantum system. The KZ analysis predicts that when a parameter of the 
Hamiltonian is quenched at a uniform rate as $t/\tau$ through the critical 
point, the density of defects in the final state shows a power-law behavior 
with the time scale of quenching $\tau$ given by 
$1/\tau^{\nu d/(\nu z +1)}$. Following recent experimental studies 
\ct{expts} of non-equilibrium dynamics of strongly correlated quantum 
systems, there is an upsurge in the theoretical investigation of related 
non-random \ct{sengupta04,zurek05,levitov06,mukherjee07} and random models 
\ct{dziarmaga06}, and models with a gapless phase \ct{polkovnikov07}. A 
generalized scaling relation for the defect density in a non-linear
quench across a quantum critical point has also been proposed \ct{sen08}.

The quenching dynamics of a spin-1/2 $XY$ chain in a transverse field, when 
either the transverse field \ct{levitov06} or the interaction \ct{mukherjee07}
is quenched from $-\infty$ to $+\infty$ at a uniform rate $t/\tau$, has been 
studied extensively in recent years, and the defect density is found to obey 
the KZ prediction. It should also be noted that recent studies of the 
two-dimensional Kitaev model indicate a generalization of the KZ prediction 
when quenched along a critical surface \ct{sengupta08}.

In this work, we investigate a situation where the transverse 
field is quenched back and forth across the quantum critical 
points from $-\infty$ to $+\infty$ and again from $+\infty$ to $-\infty$ 
and so on, with the functional form of the transverse field being given by 
$h = \pm t/\tau$. According to the Kibble-Zurek argument, the system fails to
evolve adiabatically near the quantum critical points, where the relaxation 
time is very large, which results in the production of kinks (in this 
case, oppositely oriented spins in the final state). In our notation, 
$n(l)$ corresponds to the defect density in the final state after $l$ 
passages through both the Ising quantum critical points to be defined below. 
Thus even values of $l$ signify $l$ half-periods and the transverse field is 
brought back to the initial value $h \to -\infty$; for odd values of $l$, the 
field in the final state tends to $+\infty$. Clearly, the case of $l = 1$ 
has been extensively studied earlier \ct{levitov06,mukherjee07}. 
In all the cases, the initial value of $h$ is large and negative so that 
all the spins are down, i.e., oriented antiparallel to the $z$-axis. If the 
dynamics were adiabatic for the entire span of time, one would expect
all spins to be down (up) in the final state following an even (odd) number 
of half-periods of quenching, and the defect is defined accordingly. 

\section{The quenching scheme and results}

The Hamiltonian of our model is given by 
\be H ~=~ - \frac{1}{2} ~\sum_n ~(J_x \sigma^x_n \sigma^x_{n+1} + J_y 
\sigma^y_n \sigma^y_{n+1} + h \sigma^z_n), \label{h1} \ee 
where the $\sigma$'s are Pauli spin matrices satisfying the usual commutation 
relations. The strength of the transverse field is denoted by $h$, and 
$J_x - J_y$ is the measure of anisotropy; $J_x$, $J_y$ and $h$ are all 
non-random variables. The Hamiltonian in Eq. (\ref{h1}) can be exactly 
diagonalized using the Jordan-Wigner transformation which maps a system of 
spin-1/2's to a system of spinless fermions \ct{lieb61,kogut79,bunder99}.
Diagonalizing the equivalent fermionic Hamiltonian in terms of 
Bogoliubov fermions, we arrive at an expression for the gap in the 
excitation spectrum given by \ct{lieb61,bunder99}
\be \ep_k = [h^2 +J_x^2 + J_y^2 + 2 h (J_x + J_y) \cos k + 2 J_x J_y
\cos 2k ]^{1/2}. \label{ek} \ee
The gap given in Eq. (\ref{ek}) vanishes at $h = \mp (J_x + J_y)$ for wave 
vectors $k =0$ and $\pi$ respectively, signaling a quantum phase transition 
from a ferromagnetically ordered phase to a quantum paramagnetic phase known 
as the ``Ising" transition. On the other hand, the vanishing of the gap at 
$J_x =J_y$ for $|h|<(J_x+J_y)$ at an ordering wave vector $k_0 = \cos^{-1}
(h/2J_x)$, signifies a quantum phase
transition which belongs to a different universality class from the Ising 
transitions between two ferromagnetically ordered phases. In our quenching 
scheme, the system will be swept across the Ising critical points only. 

Let us first briefly recall the case with $l=1$ \ct{levitov06,mukherjee07}.
When projected to the two-dimensional subspace spanned by the state vectors 
$\ket 0$ (empty state) and $\ket {k,-k}$ (two fermion state), the Hamiltonian 
takes the form
\ba H_k (t) &=& -~[ h ~+~ (J_x+J_y)\cos k] ~I_2 \non \\
&+& \left[ \begin{array}{cc} h+ (J_x + J_y) \cos k & i (J_x - J_y) \sin k \\
-i(J_x - J_y) \sin k & -h-(J_x+J_y)\cos k \end{array} \right], \non \ea
where $I_2$ denotes the $2 \times 2$ identity matrix. Therefore, the many-body
problem is effectively reduced to the problem of a two-level 
system, with the two levels being the states $\ket{0}$ and $\ket{k,-k}$. 
The general state vector $\psi_k(t)$ evolving in accordance with the 
Schr\"odinger equation $i \partial_t\psi_k(t) ~=~ H_k(t) ~\psi_k(t)$ can be 
represented as a linear superposition 
$\ket{\psi_k(t)} = C_{1k}(t)\ket{0} + C_{2k}(t)\ket{k,-k}$, with the initial 
condition $C_{1k}(-\infty)=1$ and $C_{2k}(-\infty)=0$. 
The off-diagonal term of the projected Hamiltonian, $\Delta= (J_x - J_y)
\sin k$, represents the interaction between the two time-dependent levels 
$\ep_{1,2}=\pm [h(t)+(J_x + J_y)\cos k]$. The Schr\"odinger equation 
given above is identical to the Landau-Zener (LZ) problem of a two-level 
system \ct{landau} and therefore, the transition probability of excitations 
at the final time is given by \ct{sei05}
$p_k= |C_{1k}(+\infty)|^{2}=~ e^{-2 \pi \ga} , \label{pk}$
where $\ga = $ $\Delta^2 /\alpha$, so that 
in the present case $p_k(l=1) = p_k(1)= e^{-\pi \tau (J_x - J_y)^2 \sin^2 k}.$
This immediately leads to an expression for the density of kinks or down 
spins $n$ in the final state
\be n(l=1) ~\equiv ~ n(1) ~=~ \int^\pi_0 ~\frac{dk}{\pi} ~p_k ~\simeq ~
\frac{1}{\pi \sqrt{\tau}~ |J_x - J_y|}, \label{n1} \ee
in the limit of large $\tau$. The $1/\sqrt{\tau}$ decay of the defect density
is in accordance with the Kibble-Zurek prediction\ct{zurek05,levitov06}.

We shall now generalize the quenching dynamics to the case $l = 2$, when the 
system is brought back to the state with $h \to -\infty$ from the final state
of the case $l=1$, with the initial condition given by $|C_{1k}(\infty)|^2 = 
e^{-2\pi\ga}$ and $|C_{2k}(\infty)|^2 = 1 - e^{-2\pi\ga}$. With a view to 
estimating the non-adiabatic transition probability, we consider two 
time-dependent states $\ket {1}$ and $\ket {2}$ with energy $\ep_1(t)$ and 
$\ep_2(t)$, respectively where $\ep_1(t) - \ep_2(t) = \al t $, $\al$ being a 
constant and the time-independent coupling between the states being $\Delta$.
Let us also assume that the time $t$ goes from $-\infty$ to $+\infty$ (the 
forward path in Fig. 1).

\begin{figure}[htb]
\includegraphics[height=1.5in,width=2.1in,angle=0]{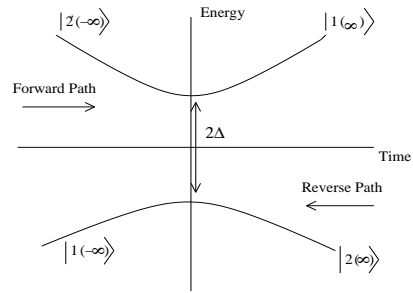}
\caption{The time-dependent energy levels of the Landau-Zener Hamiltonian.
The minimum gap is $2 \Delta$.} \end{figure}

Defining a general state vector as $\vert\Psi(t)\rangle = C_1(t)\vert1(t)
\rangle + C_2(t)\vert2(t)\rangle$, where $|C_1(t)|^2 (|C_2(t)|^2)$ is the 
probability of the state $\ket{1} (\ket {2})$ at time $t$, 
we can recast the Schr\"odinger equation in the form
\be \frac{d^2}{dz^2}\overline{U}(z) + (r + \frac{1}{2} - \frac{1}{4}z^2)
\overline{U}(z) = 0, \ee
where $z = e^{-\frac{\pi}{4}i}\al^{\frac{1}{2}}t$, $r = i\Delta^{2}/
\al$, and $U(t) = \overline{U}(z)$ is related to
$C_2(t)$ through the relation $U(t) = C_2(t)e^{i\int^t_{-\infty}{dt'
\ep_2(t')}}e^{\frac{i}{2}\int^t_{\infty}{dt'(\ep_1(t') - 
\ep_2(t'))}}$. Focusing on the special case with $|C_1(-\infty)| = 1$ 
and $|C_2(-\infty)| = 0$, we have $\overline{U}(z) = A D_{-r-1}(-iz)$ as the 
particular solution of the Weber's differential Eq. (4) \ct{sei05,whittaker58}.
We remark that the axis in the $z$ plane which corresponds to $t$, is along 
$e^{-\frac{1}{4}\pi i}$ for $t > 0$, and along $e^{\frac{3}{4}\pi i}$ for 
$t < 0$. The constant $A$ is determined from the initial condition 
$|C_1(-\infty)| = 1$ by taking the asymptotic expansion of $D_{-r-1}(-iz)$ 
along $e^{\frac{3}{4}\pi i}$ (or $t\to -\infty$ limit). This finally yields 
$D_{-r-1}(-iRe^{\frac{3}{4}\pi i}) \approx e^{-\frac{\pi}{4}(n + 1)i}
e^{-iR^2/4}R^{-n - 1}$, where $R=\sqrt {|\al|} |t|$ and we find 
$|A| = \ga^{1/2} e^{-\pi \ga /4}$. The solution in the limit $R \to +
\infty$ (or $t\to +\infty$) along $z = Re^{-\frac{1}{4}\pi i}$ leads to the
final expression $|C_1(\infty)|^2 = e^{-2\pi \ga}$ (a result already 
used for the case $l = 1$). The parameter $\ga$ = $\Delta^2 /\alpha
$ depends on the magnitude of the 
slope $\al$ of the two approaching states and the interaction $\Delta$.
Using the above results, we therefore get a recursive relation for
the probabilities after the $(l+1)$-th quenching (with $l \geq 0$) given as

\ba |C_1(-(-1)^{l+1}\infty)|^2 &=& e^{-2\pi \ga}|C_1(-(-1)^l\infty)|^2 
\non \\ &+& (1-e^{-2\pi\ga})|C_2(-(-1)^l\infty)|^2, \ea
\ba |C_2(-(-1)^{l+1}\infty)|^2 &=& (1-e^{-2\pi\ga})|C_1(-(-1)^l\infty)|^2 
\non \\ 
&+& e^{-2\pi \ga}|C_2(-(-1)^l\infty)|^2, \ea
Eqs. (5) and (6) look incomplete at first sight because they 
do not contain any cross terms like $C_1 (\pm \infty)C_2 (\pm \infty)^*$
or $C_2 (\pm \infty)C_1 (\pm \infty)^*$. However, this is because at
$t \to \pm \infty$, the coefficients $C_1 (t)$ and $C_2 (t)$ vary rapidly 
with time as $\exp [\pm (i/\hbar) \int^t dt' E (t')]$ \ct{levitov06}. Hence, 
the two cross terms given above vary rapidly with the initial time (which is 
going to $+ \infty$ or $-\infty$ depending upon the number of repetitions as 
explained above) and independently of each other for different values of $k$. 
Their contribution to the defect density therefore vanishes upon 
integration over $k$ due to the presence of terms like $\cos (T \cos k)$, 
$T$ being the time at which the terms are calculated (which is $\pm \infty$ 
in our case). On the other hand, the terms given in Eqs. (5) and (6), namely,
$| C_1 (\pm \infty) |^2$ and $| C_2 (\pm \infty) |^2$ have no such rapid 
variations, since any arbitrary large phase in $C_1 (\pm \infty)$ cancels the 
exactly opposite phase appearing in $C_1 (\pm \infty)^*$ (with a similar 
argument holding for
$C_2 (\pm \infty)$. Thus Eqs. (5) and (6) follow from an exact solution of the
Landau-Zener problem with general initial conditions $C_1 (-\infty)$ and
$C_2 (-\infty)$, once we use the fact that the phases of these initial
amplitudes are rapidly varying and are therefore uncorrelated with each
other. This is also the reason why we can consider the state obtained
after one or more quenches as a mixed state with an entropy given by
the expression in Eq. (9) below.
 
Using the above results, one finds
the probability of a non-adiabatic transition to be
\be p_k (2) =|C_{2}(-\infty)|^2 = 2e^{-2\pi\ga}(1 - e^{-2\pi\ga}). \ee
The transition probability $p_k(2)$ (Fig. 2) is maximum for 
$k= \pi/2$ for small values of $\tau$, whereas for higher values of 
$\tau$, there are two maxima symmetrically placed around $k=\pi/2$ at 
$k_0 = \sin^{-1} \sqrt \frac{\ln 2}{\pi\tau (J_x-J_y)^2}$ which gradually 
shift to $k=0$ and $\pi$ for very large $\tau$. In the limit of $\tau 
\to +\infty$, there is no non-adiabatic transition as the dynamics is 
perfectly adiabatic. The observed behavior of $p_k(2)$ also corresponds to 
the existence of an inherent time scale $\tau_0$ (for $k_0 = \pi /2$) in 
the problem which separates the regions of small and large $\tau$. 
It is interesting to note that the non-adiabatic transition probability for 
the forward path attains the minimum value\ct{levitov06} of $1/2$
at the wave vector $k=\pi/2$ for $\tau =\tau_0 = (\ln 2)/[\pi (J_x - J_y)^2]$,
while for the reverse path, although $p_{k_0=\pi/2, \tau_0}(2)$ is once again 
$1/2$, this is the maximum possible value of $p_k(2)$. 

\begin{figure}[htb]
\includegraphics[height=1.7in,width=2.9in]{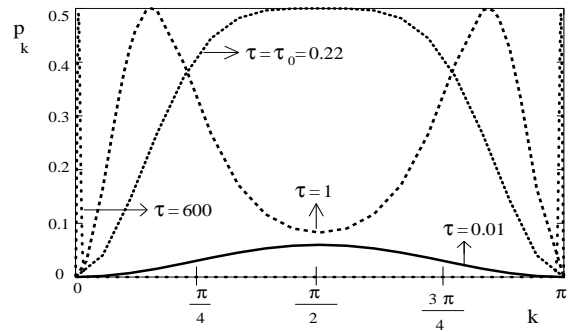}
\caption{Variation of $p_k(2)$ with $k$ for different values of $\tau$ as 
obtained analytically for $J_x - J_y = 1$.} \end{figure}

The kink density $n(2)$, i.e., the density of the up spins in 
the final state with $h \to -\infty$, is again related to $p_k(2)$ as 
$n(2) = \frac{1}{\pi} \int_0^\pi dk p_k(2)$ (Fig. 3 (a)).
In the limit of large $\tau$, we find that
\be n(2) = \frac{2}{\pi \sqrt \tau (J_x - J_y)} (1 - \frac 1 {\sqrt 2}) .\ee
Clearly, the magnitude of the defect density after a full period ($l=2$) case,
is smaller than the $l=1$ case given in Eq. (3) in the limit of large 
$\tau$, and also in the limit of small $\tau$ when the defect density is 
maximum for $l=1$. This establishes the existence of a corrective
mechanism during the reverse quenching process, arising from the fact that 
the maximum possible value of $p_k(2)$ is $1/2$, which makes the area under 
the curve of $p_k$ $vs$ $k$ smaller for the $l = 2$ case as compared to 
the $l = 1$ case. The defect density $n(2)$ (see Fig. 3)
attains a maxima around $\tau \sim 2\tau_0$; eventually there is a 
$1/\sqrt\tau$ decay in the asymptotic limit of $\tau$. We can explain the 
$n$ $vs$ $\tau$ behavior in the following way: in the limit of small 
$\tau$, the system fails to evolve appreciably, and always remains close to 
its initial state for both $l=1$ and $l=2$, whereas for large $\tau$, 
the system evolves adiabatically at all times except near the
quantum critical points. In either situation the non-adiabatic transition 
probability $p_k(2)$ and hence the density of defects $n(2)$ are small 
though it is larger in the intermediate range of $\tau$.

\begin{figure}[htb]
\includegraphics[height=2.9in,width=3.3in]{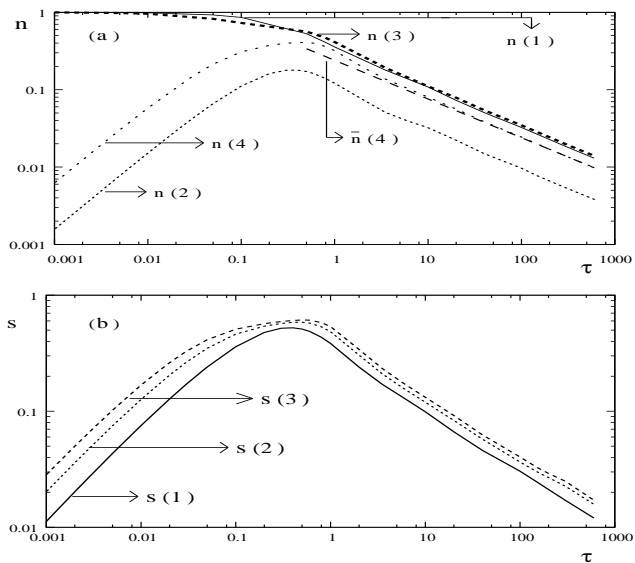}
\caption{(a) Variation of $n(1)$, $n(2)$, $n(3)$ and $n(4)$ with $\tau$, and 
(b) variation of $s(1)$, $s(2)$, and $s(3)$ with $\tau$ as obtained by 
numerically integrating Eqs. (9) and (10) with $J_x-J_y =1$. In (a), 
$\bar {n(4)}$ denotes the defect density as obtained from the analytical 
expression in Eq. (11). In the limit of large $\tau$, the numerical results 
match perfectly with the analytical results.} \end{figure}

Although the final state after a full period is a pure state, locally it may 
be viewed as a mixed state described by a decohered reduced diagonal density 
matrix with elements\ct{levitov06} $p_k$ and $1-p_k$.
The Von Neumann entropy density of the final state is given by
\ba s ~=~ -\int^\pi_0 \frac{dk}{\pi} ~[~p_k \ln (p_k) ~+~ (1 - p_k) \ln 
(1 - p_k) ~]. \ea
$s(2)$ shows a similar behavior with $\tau$ as $n(2)$ (Fig. 3(b)), and the
behavior can be justified along the same line of arguments as given above.

For an arbitrary number of repeated quenching, the non-adiabatic 
transition probability $p_k(l) = (1 - e^{-2\pi\ga}) - (1 - 2e^{-2\pi\ga})
(1 - p_k(l - 1))$. We can simplify this to obtain 
\ba p_k(l) &=& \frac{1}{2} - \frac{(1 - 2e^{-2\pi\ga})^l}{2}. \ea
Eq. (10) shows that $p_k(l)$ increases with $l$ for even values of $l$, while
for odd $l$, this variation depends on the value of $k$. However, in the 
asymptotic limit of $l$, $p_k(l)$ saturates to the value $1/2$ for all $k$ 
which implies that every spin is up or down with an equal probability in the 
final state following a large number of quenches.

Using $p_k(l)$ given above for any $l$, we find that the defect density scales
as $1/\sqrt{\tau}$ in the asymptotic limit of $\tau$ and can be put in the form
\ba n(l) &=& \frac{1}{2\pi (J_x - J_y)\sqrt\tau}\sum ^l_{w = 1} 
\frac{l!}{w!(l-w)!} \frac{2^w}{\sqrt w} (-1)^{w+1}. \ea
Interestingly, using Eqs. (10) and (11), for two successive values of $l$, we 
find that $n(l+1)<n(l)$ for small values of $\tau$ if $l$ is odd, a fact that 
once again emphasizes the corrective mechanism in the reverse path mentioned 
already. A similar result is also obtained for large $\tau$ for smaller
values of $l$, as shown in Fig. 3 (a). On
the other hand, $n(l+1)$ is always greater than $n(l)$ for even $l$. We also 
find that $n(l + 1) \to n(l)$ in the asymptotic limit of $l$ and $\tau$ for 
both even and odd values of $l$. Moreover, for any odd $l$, the kink-density 
$n(l)$ decreases monotonically with $\tau$, while for even $l$, $n(l)$ attains
a maxima around $\tau_0$; in the limit $\tau \to \infty$, $n(l)$ decays as
$1/\sqrt{\tau}$ in both cases. A close inspection of the 
variation of $n(l)$ with $\tau$ (see Fig. 3 (a)) also unravels other 
interesting aspects of the repeated quenching dynamics: (i) For even values 
of $l$, the characteristic scale $\tau^*$ at which the defect density
is maximum shifts to higher values of $\tau$ (from $\tau^* \sim 2 
\tau_0$ for $l=2$). At the same time we find that
$n(l)>n(l-2)$ for all values of $\tau$ so that it eventually saturates
to $1/2$ for all $\tau$ for asymptotic $l$. (ii) Similarly for odd $l$, 
$n(l+2)<n(l)$ for small $\tau$; for higher $\tau$,
$n(l+2)$ exceeds $n(l)$ following a crossover around $\tau \sim \tau^*$. 
Once again $n(l)$ saturates to $1/2$ for a large number of repeated quenches.
The behavior of the entropy density as a function of $\tau$ for any general 
value of $l$ is shown in Fig. 3 (b). It follows from Eq. (9) that 
for any given $\tau$, the local entropy density increases with $l$; in 
the limit ${l\to \infty}$, $p_k(l)$ tends to $1/2$ and therefore the local 
entropy density tends to its maximum value of $\ln 2$. 

\section{Conclusion}

In conclusion, we have studied the defect generation and entropy production 
in a transverse $XY$ spin-1/2 chain under repeated quenching of the 
transverse field 
between $-\infty$ and $+\infty$ . We have employed a generalized form of the
Landau-Zener transition formula along with the concept of uncorrelated initial
phases of the probability amplitudes so that the cross terms appearing in the 
recursion relation of probabilities vanish under integration over momentum.
Using the non-adiabatic transition formula thus obtained, we evaluate the 
defect density in the final state after an arbitrary number of quenches.

Our results show that the defect density satisfies 
$n(l) > n(l+1)$ for small $\tau$, 
if $l$ is odd; this points to the existence of a corrective mechanism in the 
reverse path. The results obtained by numerical integration of the transition 
probability and by using the analytical expression given in Eq. (11) match 
perfectly in the limit of large $\tau$. The entropy density, however, is found
to increase monotonically with the number of repetitions, showing that the 
local disorder of the system increases monotonically with $l$, irrespective of
the behavior of the kink density. For an odd number of repetitions, we observe
a monotonic decrease of the kink density with $\tau$, as seen previously for 
the widely studied $l = 1$ case. For even $l$, on the other hand, 
$n(l)$ grows for small $\tau$ but eventually decreases as $1/\sqrt\tau$ in 
the large $\tau$ limit, attaining a maxima at an intermediate value of 
$\tau= \tau^*$; $\tau^*$ shifts to higher values of $\tau$ with increasing
$l$. The difference in the behaviors of the defect density for an even and odd
number of repetitions originates from the fact that the system is expected to 
come back to its initial ground state after an even value of $l$ for a 
perfectly adiabatic dynamics. 

As mentioned above, the local entropy increases after each half-period of 
quenching.
% which signifies that the local structure of the final state is 
% getting more entangled. 
In the limit $l \to \infty$, the entropy $s(l)$ eventually saturates to its 
maximum possible value of $\ln 2$ for all $\tau$ while the 
non-adiabatic transition probability $p_k(l)$ approaches $1/2$ for all $k$. 
This result suggests that a spin remains up or down with the same probability,
irrespective of the applied field after a large number of quenches. Using 
Eq. (10), one can also define a characteristic number of repetitions $l^*
(k,\tau)$ so that for $l > l^* (k, \tau)$, the transition probability $p_k(l)
\sim 1/2$; it can be shown that $l^* (k,\tau)$ attains a minima at an 
intermediate value of $\tau$. We conclude by noting that the defect production
due to repeated quenching can be studied in systems of atoms trapped
in optical lattices \ct{duan03}, quantum magnets \ct{werns99} and spin-one
Bose condensates \ct{sadler06}.

\begin{center}
{\bf Acknowledgments}
\end{center}

A.D. and V.M. acknowledge Uma Divakaran for collaboration in related works, 
and B. K. Chakrabarti and Deepak Dhar for interesting discussions. D.S. thanks
DST, India for financial support under the project SR/S2/CMP-27/2006.

\end{document}